# Efficient and affordable catadioptric spectrograph designs for 4MOST and Hector

Will Saunders*

Australian Astronomical Observatory, PO Box 915, North Ryde, NSW 1670, Australia

## ABSTRACT

Spectrograph costs have become the limiting factor in multiplexed fiber-based spectroscopic instruments, because tens of millions of resolution elements (spectral x spatial) are now required. Catadioptric (Schmidt-like) designs allow faster cameras and hence reduced detector costs, and recent advances in aspheric lens production make the overall optics costs competitive with transmissive designs. Classic Schmidt designs suffer from obstruction losses caused by the detector being within the beam. A new catadioptric design puts the detector close to the spectrograph pupil, and hence largely in the shadow of the telescope top-end obstruction. The throughput is competitive with the best transmissive designs, and much better in the Blue, where it is usually most valuable. The design also has milder aspheres and is more compact than classic Schmidts, and avoids most of their operational difficulties.

The fast cameras mean that with 15micron pixels, the PSF sampling is close to the Nyquist limit; this minimises the effects of read-noise, which for sky-limited observations, far outweighs any difference in throughput. It does introduce pixellation penalties; these are investigated and found to be modest.

For 4MOST, low and high resolution designs are presented, with 300mm beams, 3 arms with f/1.3 cameras, and standard 61mm x 61mm detectors. Coverage is 380-930nm at R=5000-7000, or R~20000 in three smaller ranges. A switchable design is also presented. For Hector, a design is presented with 2 arms, 380-930nm coverage, and R=3000-4500; a 4-armed design with smaller beam-size and detectors is also presented. The designs are costed, and appear to represent excellent value.

## 1. INTRODUCTION

Over the last 20 years, multi-fiber spectroscopy has seen large increases in both fiber numbers and the required number of spectral pixels per spectrum, with each now measured in thousands, and their product in tens of millions. This means that spectrograph costs are now the limiting factor in the multiplex gain for complete multi-fiber instruments, and so there is a strong desire to find affordable spectrograph designs.

For 4MOST, there is a requirement for a minimum of ~6000 spectral resolution elements for both the High and Low Resolution Spectrographs (HRS, LRS). This can just be accommodated with 3-armed systems and standard 61mm x 61mm detectors, with a camera speed of ~f/1.3 (this assumes a fiber size of 85μm, an f/3 collimator, and that the FWHM is 1/1.15 × the projected diameter[1]). This is unfeasibly fast for transmissive designs, but is a modest speed for Schmidt or Schmidt-variant cameras. The principal drawbacks with Schmidt cameras are (a) the inaccessibility of the detector, and (b) the obstruction losses caused by it being in the beam. The 'double-pass Maksutov' design [1] partially solves both problems. It puts the detector close to the spectrograph pupil (which is as usual at or near the grating). The spectrograph pupil is an image of the telescope pupil (albeit a blurred one because of Focal Ratio Degradation), and the telescope pupil also has a large obstruction at its center, due to the telescope top end. There is then a substantial shadowing effect, and hence reduced overall obstruction losses. The reduced sensitivity to detector obstruction allows us to put the CCDs in 'lollipop' dewars such as the long neck version of the SpecInst 1100S [2], with the field-flattening lens as dewar window. Overall, the design has simple optics, excellent imaging, good efficiency, and avoids the costs, risks and operational difficulties of using Schmidt cameras as dewars. The name is a bit of a misnomer, because the corrector lenses are (mildly) aspheric, as well as having the characteristic Maksutov meniscus shape.

---

[1] The general relationship between fiber diameter, aberrations, scattered light (PSF wings) and FWHM has been investigated. A value FWHM=D/1.15 (where D is the projected fiber diameter) is within a few % of the correct value for a wide range of imaging and scattered light performance, and is assumed here throughout.
*will@aao.gov.au

When designing a low dispersion VPH spectrograph with fixed resolution and wavelength coverage requirements, beam-size has a very small effect on the VPH efficiency profile, as long as the grating angles remain in the regime 10-30° [1]. Therefore, we can envisage large beams, both to minimise the obstruction losses, and also to reduce the field angles (and hence distortion, dispersion variation, and slit curvature). For a high resolution spectrograph, a large beam is very desirable, since this reduces the grating angles, and hence increases VPH grating efficiency through better $s$ and $p$ polarisation efficiency matching [3]. A beam-size of 300mm is convenient for both LRS and HRS, in that (a) the grating angles are then ~14° for LRS (R=5000-7000) and ~40° for HRS (R~20,000), both within the VPH comfort zone; (b) it allows single-exposure VPH gratings, even at high dispersion, and (c) the cost of the other optics is not prohibitive.

On-axis collimators are used, because (a) the resulting obstruction losses are modest for such a large beam, (b) these are further reduced by the shadowing between slit and dewar, and (c) because collimator and camera aberrations can be traded if both are on-axis, greatly simplifying the design, and (d) because an off-axis f/3 collimator is a major challenge and cost driver in itself, of comparable difficulty to the cameras.

The baseline 4MOST designs are fixed format, LRS and HRS. However, it would be very desirable to have designs which are partially or fully switchable between LRS and HRS use, and these have also been investigated, and look feasible. The plan of the paper is to address the LRS and HRS designs for 4MOST first, and then the differences required for Hector.

## 2. DESIGN DETAILS

The optical layouts for LRS and HRS are shown in Figure 2. The collimator is f/3. A narrow field-lens is included (of S-FSL5, and gelled to the slitlets as in AAOmega [4]), to improve throughput and to ensure pupil-centricity of the design. To accommodate the 2 dichroics requires having the first dichroic right at the slit (with the slit unit in a slot in the dichroic substrate). DESI also has this feature [5], and VIRUS has a fold-mirror in the same place [6]. The collimator design leaves ~165mm axial distance for the slitlets and the fibers to be routed out of the beam, which should be sufficient.

Both collimator and camera corrector lenses are also S-FSL5, with aspheres on the concave side for ease of testing, and maximum aspheric slope <10μm/mm. The field-flattening lens is made of S-YGH51, as used for AAOmega [4], and also Hermes [7]. It allows a very compact lens, with small asphericity (<4μm/mm). It has ~7.5% absorption at 380nm, which could be avoided by using fused silica or CaF2, but the lens would then be much larger, causing several % additional obstruction losses at all wavelengths. Just prior to submission of this paper, concern has been raised about possible radioactivity of S-YGH51, especially when used in conjunction with bulk silicon detectors. Because of this, a sensitivity test to this glass choice has been performed, and both S-LAH66 and S-LAH64 are suitable alternatives, with negligible impact on performance.

The pupil stop is at the camera corrector for LRS, and at the grating for HRS. There is a minimum of ~50mm for the physical depth for the camera head, which should be enough, and can in any case be increased if needed without severe penalty. The camera speed was unconstrained, subject to getting the required demagnification of 0.428 (determined by the sampling and resolution requirements). The WFNO is ~f/1.31, and always a bit faster in the red. The detectors are E2V CCD231-084 (4096 × 4096, 15μm pixels) for LRS and CCD290-66 (6144 × 6144, 10μm pixels) for HRS, back-illuminated, deep-depletion, and high-rho for the blue/green/red cameras respectively. The design is physically rather large, with overall width is 1800mm, length 2600mm, and height 600mm. The weight of optics is ~300kg for each spectrograph, mostly (~200kg) for the mirrors.

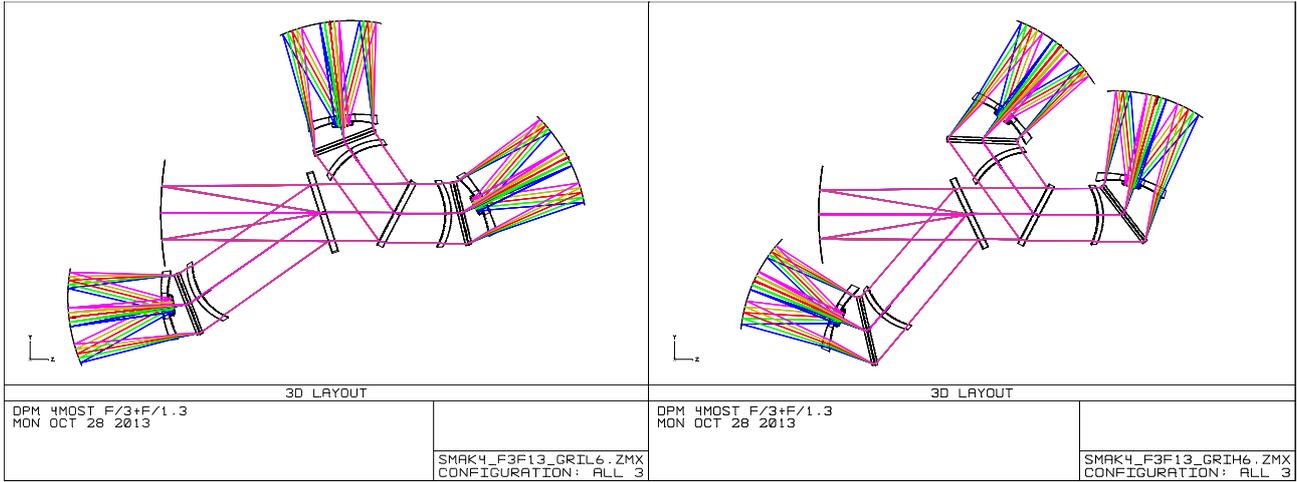

Figure 2. (a) DPM LRS and (b) HRS optical designs for 4MOST. For each arm, there is a Schmidt-like collimator, with a weak aspheric corrector lens, then the grating, then a Maksutov-like camera, but where the light passes through the corrector twice, putting the detector outside the camera entirely.

## 3. WAVELENGTH COVERAGE AND RESOLUTION.

The LRS design was optimised for a wavelength coverage of 380-930nm. The imaging performance and throughput allow the coverage to be stretched further to both the red and the UV if wanted, the limits come entirely from the resolution and sampling requirements. The predicted resolution for each camera is given in Figure 3(a). The resolution is R>5140 over the unique coverage range of each camera. For HRS, the wavelength range is $\Delta\lambda/\lambda = 0.098$ in each camera, with resolution R=19500 at the blue end, 20400 at the central wavelength, and 21400 at the red end.

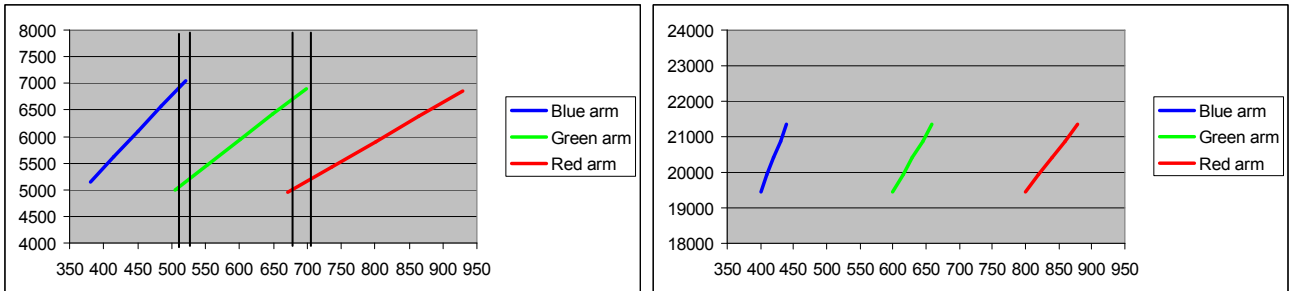

Figure 3. (a) LRS Resolution vs wavelength for the 3 cameras. The dichroic crossover ranges are also shown. (b) HRS resolution versus wavelength. The HRS wavelength ranges are still TBC and those shown are nominal.

## 4. SAMPLING AND READ NOISE

The sampling varies from 2.11 to 2.19 pix/FWHM. for the LRS, and from 3.03 to 3.44 pix/FWHM for HRS. The LRS sampling is close to the Nyquist limit, and this has implications for the resolution. When the features are weak compared with the overall continuum (as for both faint sky-limited work, and for weak stellar features), the effects can be straightforwardly quantified (Figure 4). As the Nyquist sampling limit is approached, both the centroid and the width show (a) an average loss of precision with finite sampling, and (b) a dependence of the precision on the actual pixellation of the feature. Effect (a) is very well modelled as a loss of resolution, given by adding the pixel variance to that derived from the optics. Effect (b) is negligible for measuring centroids, and only a ~1% effect for measuring line-widths. The overall equivalent resolution penalty, compared with conservative 2.5 pix/FWHM sampling, is always < 2.4%, and so we demand that the minimum resolution from the optics should be increased by this amount, to compensate.

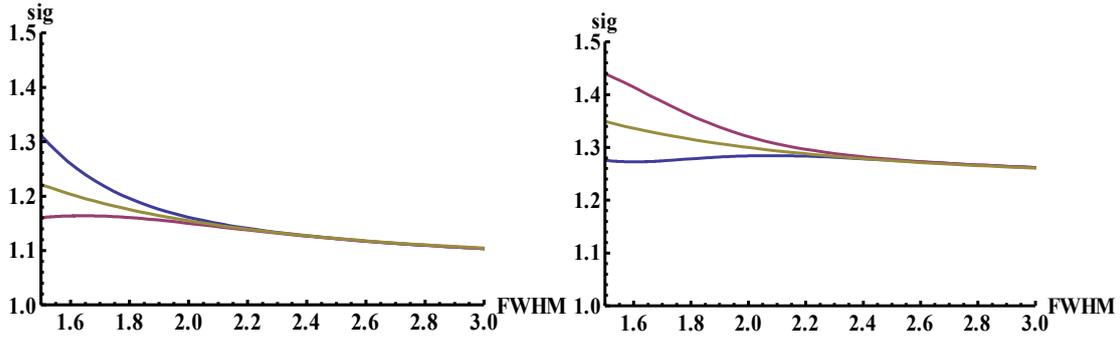

Figure 4. Sigma vs sampling for (a) the centroid and (b) the width of a pixelated Gaussian feature of small and fixed equivalent width with respect to the sky+object continuum. The x-axis shows the sampling (the FWHM in pixels), while the y-axis is the standard deviation on that centroid or width, with arbitrary overall scaling. In each case, the extreme curves show the worst-case effects of varying the pixellation, while the central line is a simple model given by adding the pixel variance to that of the Gaussian feature. Note that the best pixellation for measuring the centroid (line centered on a pixel center) is the worst for measuring the width, and vice versa.

Read Noise forms a separate very strong driver for close-to-Nyquist sampling rates. For 20-minute sky-limited LRS observations, the rate of collected photons from the sky continuum in the *peak* spatial pixel will be ~ $150\lambda/F^2$, where $\lambda$ is the wavelength in microns, $F$ is the sampling (i.e. the FWHM in pixels), at all wavelengths, so at best some tens of counts. The dependence of the survey efficiency on Read Noise and peak counts [1], can be very remarkably well approximated as $\eta = (1+ 3/2\ (R^2/N)^{9/5})$, where $\eta$ is the effective efficiency (the efficiency giving the same S/N in the absence of Read Noise), $R$ is the read noise and $N$ the counts in the peak spatial pixel (see Figure 5). Even with state-of-the art detectors and fast cameras, the resulting efficiency penalties are large, and their variation is at least as important as any plausible variations in throughput between different designs (Table 1).

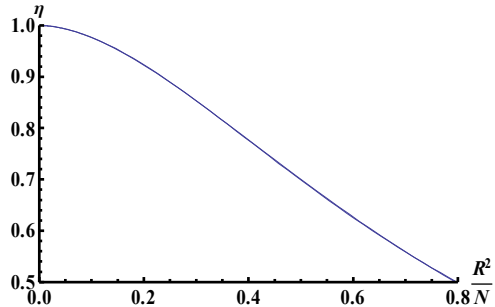

Figure 5. Efficiency penalty as a function of the ratio of read noise variance to counts in the peak spatial pixel, for a Gaussian profile, together with the analytic approximation $\eta = (1+ 3/2\ (R^2/N)^{9/5})$.

Table 1. Read-noise efficiency penalty (the equivalent throughput giving the same S/N in the absence of read noise) as a function of read noise sampling, for 20 minute sky-limited 4MOST observations. The additional efficiency penalty from read-out *time* is not included.

| | Blue (400nm) | | | Green (600nm) | | | Red (900nm) | | |
|---|---|---|---|---|---|---|---|---|---|
| FWHM | 3.0 pix | 2.5 pix | 2.0 pix | 3.0 pix | 2.5 pix | 2.0 pix | 3.0 pix | 2.5 pix | 2.0 pix |
| RN/pixel | | | | | | | | | |
| 2.5 e- | 0.424 | 0.591 | 0.764 | 0.609 | 0.751 | 0.870 | 0.764 | 0.861 | 0.932 |
| 2.0 e- | 0.627 | 0.764 | 0.878 | 0.777 | 0.870 | 0.937 | 0.878 | 0.932 | 0.968 |
| 1.5 e- | 0.825 | 0.900 | 0.952 | 0.907 | 0.949 | 0.976 | 0.952 | 0.975 | 0.988 |
| 1.0 e- | 0.952 | 0.975 | 0.988 | 0.976 | 0.988 | 0.994 | 0.988 | 0.994 | 0.997 |

## 5. INPUT BEAM APODISATION

The far-field beam from the 4MOST fibers will be apodised, by an amount that depends on the fiber position in the focal plane, the spine tilt, and on FRD. The beam is further altered by obscuration from the slit unit and the dewar. The 4MOST tilt distribution has been investigated [8], with the results (a) using the rms tilt gives an excellent measure of the average collimator and obstruction losses, and that (b) the rms tilt will be ~1.4°. A new model for FRD has been devised [9] which appears to reproduce the dependence of FRD on speed [10] extremely well. This model was used to calculate the far-field beam from the fibers, as a function of tilt and FRD quality, and the results are shown in Figure 6.

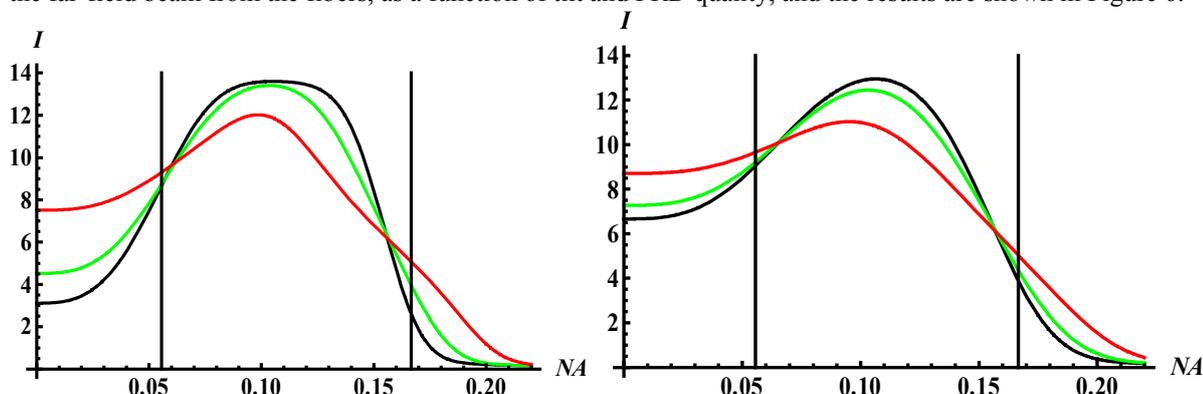

Figure 6. Radial far-field profiles for the 4MOST beam exiting entering the spectrograph, for no spine tilt (black), 1.25° tilt (green) and 2.5° tilt (black). Actual rms tilt is 1.4° and maximum tilt 2.4°. The size of the dwar head radius and the collimator acceptance speed are shown as the vertical bars. (a) is for mediocre FRD (1.85° annulus FWHM for f/3.24 collimated input), (b) is for terrible FRD (3.0° FWHM).

The resulting apodisation factors were fed to ZEMAX, for 0°, 1.25°, 2.5°. Obstructions were also included for the slit unit, dewar head, and dewar neck. The rms radii were 5-10% better than the results without any apodisation, and showed only a few % variation between different tilts. So for this design at least, the impact of varying spine tilt is modest, certainly much less than the scattering effects discussed below.

The effect of spine tilt on the overall PSF shape has also been studied [11]. It was found that the different spine tilts introduced variations in the PSF of order 1-2% of its peak value. That is large enough to require that calibration frames be taken with the same spine tilts as the data, but again negligible in terms of resolution or cross-talk.

## 6. IMAGING PERFORMANCE

The image quality (neglecting apodisation) is <7.2μm rms radius for LRS and <6.3μm for HRS, in all colors and slit position, with typical values are 4-5μm. This is slightly better than AAOmega, for which the image quality was, in retrospect, better than was required (because sky subtraction methods now always incorporate corrections for PSF variation). The spots are shown, compared with the projected fiber size, for all configurations for both LRS and HRS, in Figure 7.

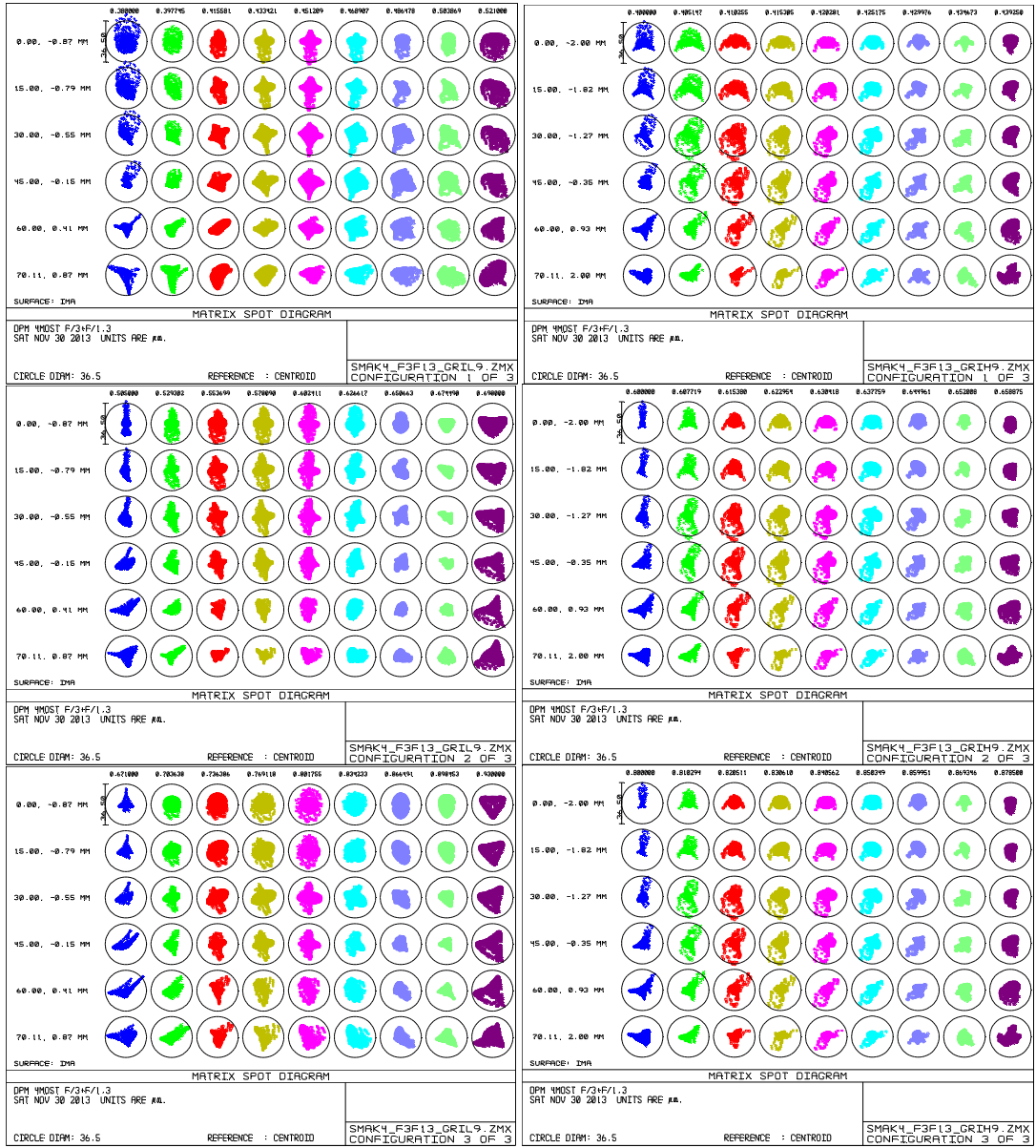

Figure 7. Top: LRS (left) and HRS (right) spot diagrams for Blue arm. Center: ditto for Green arm. Bottom: ditto for Red arm. In all cases the circle is 36.5μm diameter (the projected fiber size). The ray-tracing overestimates the true aberrations, because no account is taken of the input beam apodisation. This gives a 5-10% reduction in the rms spot size.

These aberrations are small compared with the fiber size. Figure 8(c) shows the full aberrated image for one of the worst images. The box size is 75μm, which is the smallest conceivable separation between spectra (see below). The image is wholly contained within the box, and would leave a full pixel gap between spectra. So the cross-talk arising from the geometric optics is negligible for this design.

Because the beam is so large compared with the detector size, the field-angles are modest, so there is not much distortion, slit curvature or variation of the dispersion, allowing very efficient CCD usage. The slit has been curved in the spectral direction, by just 1.8mm, to ensure that for LRS use, the light at the limiting wavelengths for each camera is on the detector for all fibers. The same value has been assumed for HRS use, though a larger value (the optimal value is ~4mm) could be used if the slit unit design allowed it.

## 7. DIFFRACTION EFFECTS

All the image qualities shown so far take no account of diffraction effects. The design presented here should not have strong differences in the diffraction pattern across the detector, because the obstruction is at the pupil. Figure 8 shows the PSF including diffraction at the center and one corner of the detector, showing that (a) the diffraction artefacts are at a level at least 2 orders of magnitude below the peak, (b) they are contained entirely within 1-2 pixels of the core, (c) they do not degrade the geometric aberrations unreasonably, and (b) they are invisible in the integrated PSF. Note that the spatial direction is horizontal in Figure 8, so the diffraction spikes affect the spectral PSF, but not the spatial.

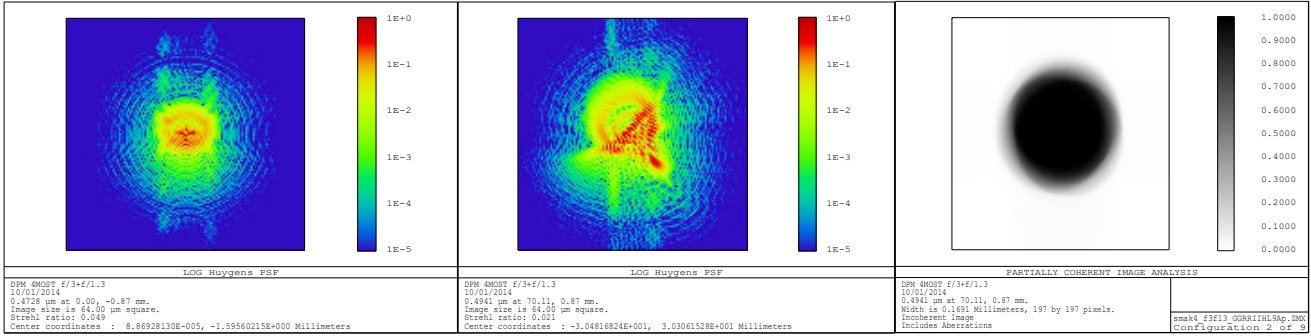

Figure 8(a,b). Aberrations, including diffraction, for field at the center and corner of the detector. Box size is 64μm.
Figure 8 (c) The resulting PSF in the corner of the detector. Box is 75μm.

## 8. OBSTRUCTION LOSSES AND THROUGHPUT

Any on-axis design suffers from obstruction losses. These depend not only on the spectrograph design, but also on the far-field apodisation pattern coming from the fibers, which only partially preserves the central obstruction from M2, and the preservation is made worse by Echidna spine tilt [11]. The resulting additional spectrograph obstruction losses have been calculated directly as a fraction of the light within f/3, and including the shadowing between M2, slit, and camera. For LRS, this overlap is very great, because the pupil is so close to the detector (so the detector sits close to the middle of the top-end shadow for all fibers). For HRS, the shadowing is less, and also wavelength-dependent, but still very helpful. The combined losses depend on the FRD performance as well as wavelength and spectrograph design, but are always less than the average collimator overfilling losses calculated from the same model. Both are summarised in Table 2.

Table 2. Collimator overfilling and spectrograph obstruction losses for different FRD performance[2].

| FRD (FWHM @ f/3.24) | Average f/3 Collimator overfilling loss | LRS | HRS best | HRS worst |
|---|---|---|---|---|
| 1.7° | 12.2% | 10.1% | 10.6% | 11.8% |
| 3.0° | 15.3% | 12.6% | 12.9% | 13.5% |

Throughput has been calculated, and is shown in Figure 9. Two values have been calculated: (a) estimates for optics+detector, and (b) 'AIP throughputs', which are for optics only and use standardised values for coatings and detectors, for comparison between designs. The AIP peak throughput is ~65% in each arm, the minima are ~57% at 380nm and ~690nm, and ~59% at 515nm and 930nm. The average throughput is ~62%. These values exceed the 4MOST throughput goals of 50% minimum and 55% average.

---

[2] All these numbers were calculated assumed 300mm spines rather than 250mm, so are too optimistic, but only by <<1%.

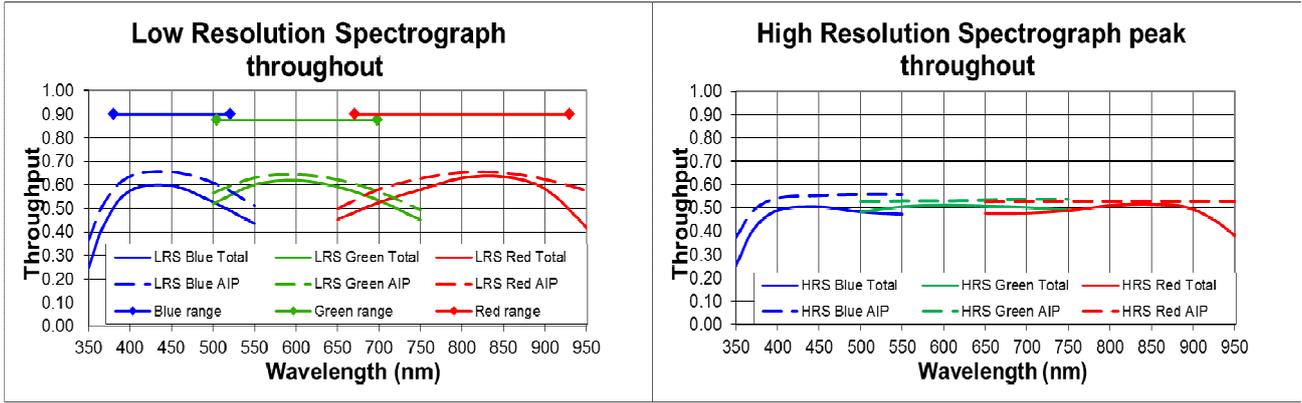

Figure 9. (a) Throughput for LRS design. 'Total' means optics+detector, and 'AIP' means optics-only, assuming AIP values where appropriate. Wavelength coverages are also shown for each arm. Note the different nomenclature for the three arms (Blue/Red/Far-Red vs Blue/Green/Red). (b) *Peak* throughput for HRS design, when the VPH grating is tuned for the wavelength shown. The minimum throughput (at wavelengths ±4.9%) is ~91% of the value shown.

For the HRS design, the modest grating angles (~40°) give excellent *s+p* joint VPH efficiency. The theoretical VPH efficiency peak is 76%, with an efficiency exceeding 70% over the HRS bandwidth of $\Delta\lambda/\lambda = 0.098$, and an average of 74%. HRS wavelength ranges are still under review, but overall throughput can be determined for any setup from this and Figures 8(b). This shows the peak throughput, for any chosen central wavelength. The minimum throughput, at wavelengths ±4.9% larger and smaller, will be ~91% of the value shown in Figure 8(b). The minimum throughput is always >48%, and the average is always >50%, in each arm, exceeding the goals of 35% minimum and 50% average.

## 9. SHUTTERS AND BACK-ILLUMINATION

The dichroic at the slit, the large beam-size, and the small pupil relief raise issues for the system shutter and the back-illumination. AAOmega uses a simple flag shutter in front of the slit unit. The proposal is to mount a similar, but even smaller, flag shutter on the slit unit itself. The shutter would be controlled by miniature rotary solenoids, e.g. that made by Takano [12]. The shutter operating time would be a few tens of msec, giving photometric uniformity <1% for both near and far field, for exposures of a few seconds or more.

It would be very attractive if the back-illumination (BI) could use this shutter also. There is a requirement to back-illuminate simultaneously with reading out, requiring a very high level of light-sealing. The proposal is to minimise the amount of BI light required, and ensure as little of it as possible is scattered. The entire slit unit would form a light-tight box, with all internal surfaces black where possible, and the shutter would shut against seals. LEDs for back-illumination would sit at the rear of the slit unit facing forward, shining over the tops of the slitlets, and be reflected into the fibers by a narrow strip of retro-reflective tape (e.g. 3M 8711) mounted on the shutter. The back-illumination is then at a very slow f-ratio (which is what the metrology camera needs), and the use of the tape reduces the required level of illumination by orders of magnitude. A micro-channel block (or smartphone privacy screen material) could be used to suppress stray light. A narrow roller-shutter (like a tape measure) would come from the top and/or bottom of the slit unit, to provide up to 3 baffled shutters in total, to give comprehensive light-sealing. The roller shutter operation does not any need great speed, because the expected read-out time is longer than required for fiber repositioning.

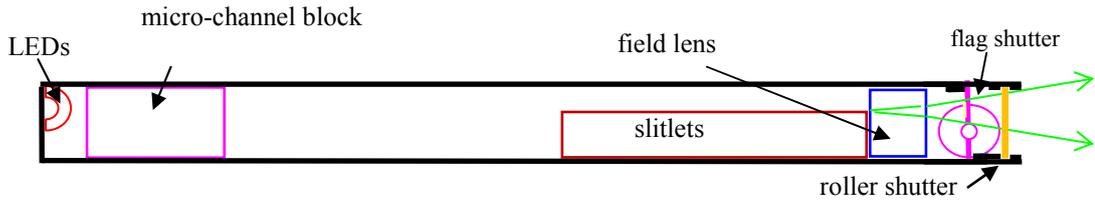

Figure 10. Schematic view from above for the proposed shutter and back-illumination. The flag shutter forms a seal against the box containing the slit-unit, and the roller shutter (coming from the top or bottom of the slit unit) ensures it is light-tight. Note the roller shutter is always open when the flag shutter is open. The back-illumination is off a very narrow retro-reflective strip (shown in yellow) on the flag shutter.

## 10. SCATTERING, CROSS-TALK AND FIBER NUMBERS

The maximum number of fibers per spectrograph depends on the cross-talk between spectra. The geometric and diffractive image analyses imply negligible cross-talk even for 5 pixel spacing between spectra, but this assumes perfect optics. Lorentzian-type weak-scattering wings on the PSF have not been included, and are clearly the dominant cause of cross-talk. The proposed 4MOST cross-talk spec is that 95% of the light be contained within $\pm\Delta/2$, where $\Delta$ is the pitch between fibers.

It happens that AAOmega + SPIRAL IFU data has similar spectrograph design [4], image quality, sampling (2.1pix/FWHM), and grating types and detectors. It can thus be used for estimating 4MOST cross-talk. This is a conservative approach, since scattering was not a primary design driver for AAOmega, so wavefront errors, surface roughnesses and grating quality are surely worse than will be specified for 4MOST. The LR gratings for 4MOST are intermediate in dispersion between the low and medium dispersion AAOmega gratings, but the HR gratings are almost identical to those for AAOmega. The SPIRAL fibers are on a pitch of 3.5 pixels, in banks of 32 fibers, with large gaps between banks of spectra (to allow Nod&Shuffle). This means that the light between $\Delta/2$ and $\Delta/2 + 3.5$pix beyond the last fiber of a bank is equal to the total light in one PSF wing for one fiber (Figure 11). We can then easily find the scattered light fraction as a function of $\Delta$, though only for particular values, including $\Delta$ = 4.5 and 6.5 pixels, which brackets the plausible pitches. The 95% is taken to exclude light that is so scattered as to form a DC offset on the detector, since the contribution from this has no spectral signature.

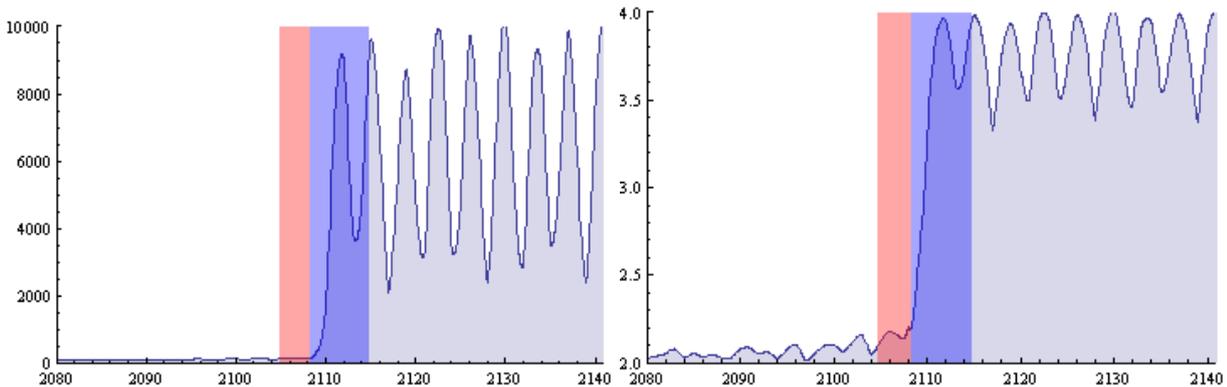

Figure 11. Example spatial profile for SPIRAL+AAOmega data with (a) linear and (b) log scaling. The red stripe has width equal to the actual pitch between fibers, and contains scattered light contributions from all the fibers. When the fibers are used at a different pitch, represented by the blue stripe, then (assuming the fibers are all equally illuminated) the total scattered light for a single fiber is equal to twice the light in the red stripe. This gives a very simple and robust measure of this scattered light.

The resulting scattered light fraction has been calculated for all the AAOmega gratings for which data is available in the AAT data archive, and is shown in Table 3. There is some dependence on colour as expected, but also 2 of the gratings (580V and 1000R) are much poorer than the others. 580V and also 385R are for much a lower grating angle than intended for 4MOST, and are not very relevant to 4MOST. 1000R must have had some problem in the manufacture, since the 1500V grating (at similar grating angle) is so good. All the other gratings with SPIRAL data meet the spec, even on a 4.5pixel pitch. There is no SPIRAL data for 1700B and 3200B; but from AAOmega+2dF data, it's clear that 3200B would easily meet the spec, while 1700B would meet easily the spec at a pitch of 6.5 pixels, but fail at 4.5 pixels. Since 3200B and 1500V meet the spec so easily, it seems 1700B could also have been manufactured cleaner. Therefore, as long as an appropriate scattered light specification is included in the grating procurement, it seems that a pitch of 5.0 pixels should allow the cross-talk spec to be met. This would allow 800 fibers to be accommodated on each spectrograph.

Table 3. Fraction of light in the PSF falling outside $\pm\Delta/2$ for a pitch $\Delta$, for various AAOmega gratings. Note that 1700D is a Dickson grating for use at 860nm, and is much thicker than the other high resolution gratings.

| Grating | 580V | 385R | 1500V | 1000R | 2500V | 2000R | 1700D |
|---|---|---|---|---|---|---|---|
| Grating angle | 8° | 8° | 22.5° | 20° | 40° | 40° | 47° |
| $\Delta$ = 4.5 pix | 8.2% | 3.8% | 2.0% | 6.8% | 1.8% | 4.0% | 4.2% |
| $\Delta$ = 6.5 pix | 6.6% | 2.4% | 1.8% | 5.5% | 1.2% | 3.2% | 3.2% |

## 11. SWITCHABLE DESIGNS

Switchable designs (between LRS and HRS use) would be very attractive, because they allow increased greater survey efficiency. Two possibilities are considered: either (a) at high resolution, allowing grating changes at fixed grating angles (and hence fixed resolution), or (b) allowing a change between HRS and LRS modes, changing both the grating and the grating angles. For (a), singlet corrector lenses remain viable. In particular, the Blue arm can be switched from ~400-440nm, to ~450-495nm, with just a refocus. The image quality is almost unchanged from the results for the fixed HRS design above. For (b), doublet camera correctors are required, to maintain the image quality (because the imaging has to now be much more achromatic). LF5 and S-FSL5 (as used in AAOmega [4]) are a suitable combination, with CTE's similar enough to allow a glued doublet. The interface does not have to be aspheric, so the additional lens is just spherical, and the additional cost should be small. However, there is a UV throughput penalty, of a few % at 380nm.
Note that it would be possible to have, for example, the Blue and Green HRS cameras of fixed format, while the Red camera could be changeable between HRS and LRS use.

The main motivation for switchable designs is driven by the Halo stars forming the only HRS targets at high Galactic latitudes: there are not enough of them to fill the HRS fibers, and their optimal Blue wavelength range is different to the other HRS cases. Both problems could be solved if the HRS Blue arm was switchable between (e.g.) 400-440nm and 450-495nm, while the HRS Red arm was switchable between HR and LR use, and the HRS Green arm remains fixed. The idea would be to observe all available Halo stars with optimal Blue and Green arm HR wavelength coverage, and put the other fibers on targets such as ELGs requiring only Red arm LRS coverage. A nice benefit of this is maintaining the singlet in the Blue arm, and hence the excellent UV/blue throughput just where it is most needed.

## 12. A 2-ARMED DESIGN FOR HECTOR

Hector is a proposed multi-IFU system for the AAT [13], using ~100 deployable Starbugs [14] on a new 3° FOV WFC [15]. About 6000 fibers are envisaged, but they can be close-packed on the detectors. The design is strongly limited by spectrograph costs. The survey is focused on specific lines (from OII 3727Å to CaII triplet) at modest redshifts ($z$<0.15). There is a strong desire to tailor the resolution to specific lines, both because of the physics (spatially-resolved emission lines can be remarkably narrow), and because of read-noise limits in the blue. The ideal resolution is R=2500 for Blue absorption line, R=4000 for the emission lines, and R=5000 for CaII triplet.

Fast f/1.3 cameras are again a good match to the proposed 100μm fibers[3], giving 2.14 pix/FWHM sampling. A two-armed system (370-590nm+570-930nm) with 4Kx4K detectors has fewer resolution elements than would be ideal, but is close enough to consider. The image quality is not quite as good as for 4MOST (because of the larger wavelength ranges), with rms spot size <7μm everywhere.

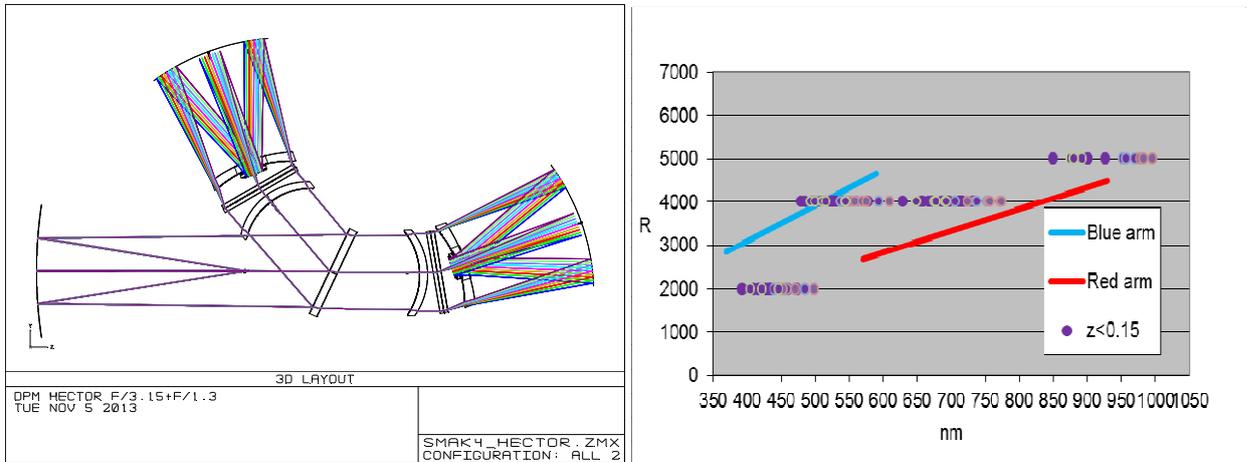

Figure 12 (a). Layout for the 2-armed Hector design, optical description as for Figure 2.
Figure 12 (b). Resolution achieved, showing also the lines of interest, over the range of expected redshifts, and the desired resolution for each.

The wavelength coverage for each arm ($\lambda_{max}/\lambda_{min} \sim 1.6$) is a big stretch for the VPH gratings. This affects both peak efficiency (since the VPH design must be tuned for width rather than peak), and that at each end of each arm. The grating angles are only ~9.5°, a bit below the VPH optimal zone. The overall peak efficiency is only ~60% (Figure 13), and the minimum efficiency is only 45% (at ~370nm, ~580nm, and >920nm). However, the peak efficiency is achieved in the crucial 400-500nm region, and it is this region that determines minimum S/N and integration times. So this is a viable and very simple design for Hector. If the fibers are butted on the slit, about 1200 can be accommodated in each spectrograph, and 5 spectrographs would be required in total.

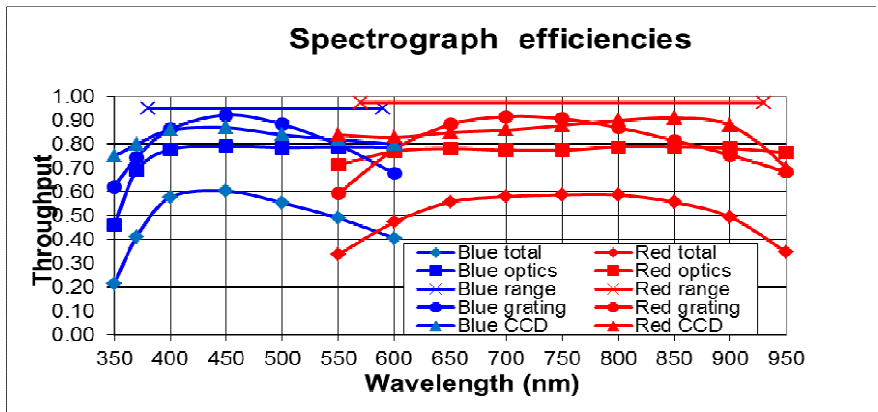

Figure 13. Predicted throughput for the 2-armed Hector design.

The simplest way to improve efficiency would be to cut off the blue or red end of the wavelength range, by excluding either OII and Ca H&K, or the Ca II triplet. Either would then allow an efficient two-armed design, with good resolution.

---

[3] Hector may include 50μm fibers also (for mapping the inner parts of galaxies). This designs are much too fast for such small fibers, and a suitable design is presented elsewhere [12].

## 13. A SMALL-BEAM 4-ARMED DESIGN FOR HECTOR

An intriguing design variant is to have a 4-armed version of the same design, with 150mm beam and 2Kx2K detectors. Amazingly, the collimator layout is straightforward (Figure 13), though it requires a dichroic at the slit, like the designs for 4MOST. The efficiency would be improved, because there is a large gain (at least 5% peak and 10% in the wings) from using peakier gratings at higher dispersion over smaller individual wavelength ranges. There's a similar but additional gain in the red, from having two flavours of red CCD (deep-depletion + high-rho). There's a hit from larger obstruction losses and the extra dichroic, but these would be smaller. Also, the image quality is greatly improved, the resolution can be remarkably well tuned to the nominal requirements (Figure 14 (b)), and Ca II triplet can be observed to the full $z$=0.15 redshift limit.

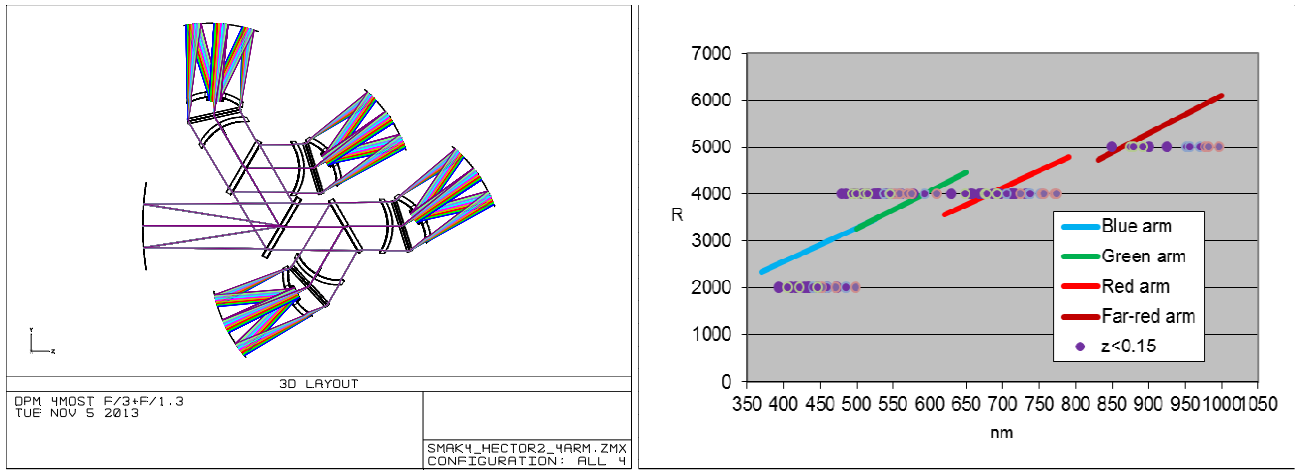

Figure 14 (a). Layout for the 4-armed Hector design, optical description as for Figure 2.
Figure 14 (b). Resolution achieved, showing also the lines of interest, over the range of expected redshifts, and the desired resolution in each case.

E2V 231-series detectors are available in 2Kx2K format. They are not yet in the required packaging, but E2V say this could be developed at reasonable cost. But this design means there will be 4 times as many detectors, dewars, lenses etc, as for the 2-armed design, for a total of 10 spectrographs and 40 dewars and detectors.

## 14. COST

ROM costings have been obtained for the major items (optics, gratings, detectors, dewars, controllers), and have been scaled from AAOmega costs for the minor ones. For 4MOST, some costings are as suggested by AIP, and AIP have also specified a hardware cost for mechanical components.

For 4MOST, the ROM costing of all hardware is ~US$1.5M for the LRS design. The cost for HRS version is virtually the same, except for the additional detector cost (with more pixels in the same format). Some of these costs (mechanical hardware, dewars, controllers) are ESO standard costs.

For the Hector 2-armed design, the costs are much less, partly because there are 2 instead of 3 cameras, and partly because off-the-shelf dewars+controllers can be used. The identified costs for optics and detectors only (so not including any mechanical hardware or labor), are US$650K per spectrograph.

## ACKNOWLEDGEMENTS

I'm extremely grateful to many people for discussions that have helped shaped these designs. They include Paul Jorden at E2V, John Kong at Precision Asphere, Kevin Toerne at SpecInst, Roger Haynes, Dionne Haynes, Roelof deJong, Sam Barden, Olivier Schnurr and Svend Bauer at AIP, Scott Croom at USyd, Graham Murray at Durham, and Jon Lawrence, Robert Content, Peter Gillingham, Greg Smith, Simon Ellis and Jurek Brzeski at AAO.